\documentclass[reprint, preprintnumbers, aps, prl, floatfix, superscriptaddress]{revtex4-2}

\usepackage{amssymb, amsmath, physics, graphicx, xcolor, orcidlink, fontawesome, hyperref}

\newcommand{\IoA}{Institute of Astronomy, University of Cambridge, Madingley Road, Cambridge, CB3 0HA, UK}
\newcommand{\KICC}{Kavli Institute for Cosmology, University of Cambridge, Madingley Road, Cambridge, CB3 0HA, UK}
\newcommand{\DAMTP}{Department of Applied Mathematics and Theoretical Physics, Centre for Mathematical Sciences, University of Cambridge, Wilberforce Road, Cambridge, CB3 0WA, UK}

\begin{document}

\title{Modeling Direct Waves in Binary Black Hole Ringdowns}

\author{Richard Dyer\,\orcidlink{0009-0008-3720-6092}}
\email{richard.dyer@ast.cam.ac.uk}
\affiliation{\IoA}

\author{Adrian Ka-Wai Chung\,\orcidlink{0000-0003-2020-3254}}
\affiliation{\DAMTP}

\author{Christopher J.\ Moore\,\orcidlink{0000-0002-2527-0213}}
\affiliation{\IoA} \affiliation{\DAMTP} \affiliation{\KICC}

\date{\today}

\begin{abstract}
    Direct waves, prompt signals propagating from a plunging object to the observer, exist alongside quasinormal modes in binary black hole ringdown. 
    It has been suggested that the properties of the direct wave are related to the event horizon; this simplifies modeling the direct wave and suggests the possibility of using it as a new observational probe of the horizon geometry.
    This paradigm is tested by extracting direct waves from numerical-relativity waveforms, adapting techniques originally developed for studying quasinormal modes.
    The direct wave is identified over a range of ringdown start times, demonstrating the utility of the horizon mode model.
    However, the direct wave frequency is found to deviate from the horizon value, limiting its utility as a probe of the event horizon.
\end{abstract}

\maketitle

{\textit{\textbf{Introduction}}---}Following a binary black hole (BH) merger, the remnant undergoes a process known as ringdown as it settles into its final state. 
Gravitational-wave (GW) observations of the ringdown provide the means to measure properties of merger remnants, test general relativity in strong fields \cite{LIGOScientific:2019fpa, LIGOScientific:2020tif, LIGOScientific:2021sio, LIGOScientific:2026wpt}, probe the Kerr nature of astrophysical BHs \cite{2019PhRvL.123k1102I, 2023PhRvL.131v1402C}, and observationally test foundational ideas such as the BH horizon area law \cite{Isi:2020tac, LIGOScientific:2025rid}.

Insight into the structure of the ringdown for a general two-body merger can be gained from the large-mass-ratio limit. 
In this regime, BH perturbation theory treats the smaller object as a test particle moving in the fixed background spacetime of the larger BH \cite{Mino:2008at, Zimmerman:2011dx}; effectively a one-body model for the ringdown.
The resulting GW perturbations satisfy a sourced wave equation, which can be formally solved using a Green's function.

The signal can be expressed as a contour integral involving the Green's function \cite{Berti:2006wq, DellaRocca:2025zbe}. 
Different parts of the contour make distinct contributions to the waveform. 
Most of the ringdown is dominated by quasinormal modes (QNMs), exponentially damped oscillations associated with poles of the Green's function \cite{Motohashi:2026mbn, DellaRocca:2025zbe}. 
At late times, the signal decays to reveal the power-law tails arising from a branch-cut contribution to the integral. 
Finally, there is the direct wave (DW) \cite{2025arXiv250909165O, 2025arXiv251001001L, Ma:2026qbq}. 
Physically, this is the signal that propagates directly from the source to the observer; mathematically, it is associated with the two quarter-circular arcs of the contour.

Recently, there has been increased interest in the DW as it is expected to improve the early-time modeling of ringdown \cite{2025arXiv250909165O, 2025arXiv251001001L, Kankani:2026byb, Ma:2026qbq, 2026arXiv260606592K}. 
Incorporating the DW might allow ringdown analyses to start earlier, increase the available signal-to-noise ratio and improve ringdown tests.

The DW is sourced by a plunging particle \cite{2025arXiv250909165O, 2025arXiv251001001L,  DellaRocca:2025zbe, Ma:2026qbq}. 
It depends not only on the background geometry but also on the particle trajectory, making it harder to model than QNMs.
Nevertheless, as the particle approaches the event horizon, frame dragging causes its angular velocity to approach the horizon frequency, $\Omega_{\rm H}$, while its radiation is exponentially redshifted for distant observers at a rate set by the surface gravity, $\kappa$.
This picture is nicely illustrated in Fig.~1 of Ref.~\cite{2025arXiv251001001L}.
Therefore, it has been suggested \cite{2025arXiv251001001L, 2026arXiv260606592K} that the DW can be modeled as a single exponential with a horizon-mode frequency 
\begin{align} \label{eq:DW_freq_beta}
    \omega^\beta_{\mathrm{DW}} = m\Omega_{\rm H} - i\kappa .
\end{align}
Both $\Omega_{\rm H}$ and $\kappa$ are geometric quantities associated with the horizon; for a Kerr BH, they depend on the mass and spin (see Eq.~\ref{eq:OmegaH_kappa}), consistent with the no-hair theorem.

If this picture is correct, then it makes the DW particularly interesting because it may directly probe geometric quantities associated with the event horizon. 
This contrasts with QNMs, whose frequencies are determined primarily by the geometry near the unstable light ring (or photon sphere). 
Consequently, DWs may offer complementary information about the BH spacetime \cite{2025arXiv251001001L}.

While this picture is physically well motivated, it remains unclear how well it describes the ringdown signals produced by the approximately equal-mass binary BH mergers now routinely observed by LIGO and Virgo \cite{2026arXiv260527223T}. 
This Letter addresses two related questions:
\begin{enumerate}
    \item How accurately, and for what range of start times, can the DW be modeled by a damped sinusoid with frequency given by the horizon quantities in Eq.~\ref{eq:DW_freq_beta}?
    \item To what extent does the horizon-mode picture derived from BH perturbation theory remain valid for the comparable-mass mergers?
\end{enumerate}

These questions are answered using techniques originally developed for studies of QNMs in numerical-relativity (NR) waveforms. 
Models incorporating both QNM and DW contributions are fit to multiple spherical harmonics of state-of-the-art Cauchy Characteristic Evolution (CCE) waveforms from the public catalog \cite{SXS_CCE_catalog}.
The performance is quantified using both simple waveform mismatches and the Bayesian model-selection techniques of Refs.~\cite{2026PhRvL.136s1403D, 2026PhRvD.113j4031D}. 
Free-frequency fits and amplitude-stability tests are also used to assess the validity of the frequency and decay rate of the DW predicted by Eq.~\ref{eq:DW_freq_beta}. 
The use of rational filters \cite{Lu:2025mwp, Ma:2022wpv} to analyze the DW is also investigated.
The DW is successfully identified for a range of ringdown start times comparable to that of the second QNM overtone and with an amplitude approximately a factor of ten below the fundamental QNM. 
However, we find that the DW frequency is only approximately given by the horizon mode in Eq.~\ref{eq:DW_freq_beta}, calling into question its ability to probe the event horizon.

{\textit{\textbf{Modeling Direct Waves}}---}The GW waveform ${h=h_+-ih_\times}$ can be expanded in terms of the (spin-weight~$-2$) spherical harmonics ${}_{-2}Y_{\beta}$,
\begin{align}
    r\mathfrak{h}(t,\theta,\phi) = M \sum_\beta \mathfrak{h}^\beta(t) {}_{-2}Y_{\beta}(\theta,\phi) ,
\end{align}
where $r$ is the distance from the BH, $M$ is the mass scale, $\mathfrak{h}^\beta(t)$ are the waveform harmonics computed from NR, and the indices $\beta=(\ell,m)$ span $\ell\geq 2$ and $-\ell\leq m\leq \ell$. 

\begin{figure}[t]
    \includegraphics[width=0.99\linewidth]{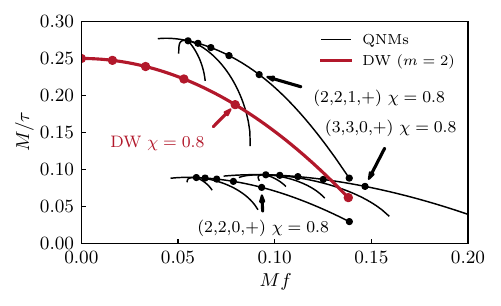}
    \caption{ \label{fig:DW_QNM_frequency_comparison} 
        Comparison of the dimensionless frequencies and damping times of the DW horizon mode and selected QNMs.
        Each mode is plotted parametrically as a line for $0\leq\chi< 1$; on selected lines dots indicate $\chi=0,\, 0.2,\, 0.4,\, 0.6,\, 0.8,\, 0.99$.
        The QNMs $\alpha=(2,m,0,+)$, $(2,m,1,+)$, and $(3,m,0,+)$ are shown for all $m$.
        Based on a similar figure in Ref.~\cite{finch_thesis}.
    }
\end{figure}

The ringdown can be modeled using a sum of QNMs,
\begin{align} \label{eq:h_QNM}
    rh_{\rm QNM}(t,\theta,\phi) = M \sum_{\alpha} C_\alpha e^{-i\omega_\alpha(t-t_0)} {}_{-2}S_{\alpha}(\theta,\phi) .
\end{align}
QNMs are indexed by $\alpha=(\ell,m,n,p)$ where $n\geq 0$ is the overtone number and $p\in\{+,-\}$ distinguishes prograde and retrograde modes.
Each QNM has a free (complex) amplitude $C_\alpha$ and a frequency $\omega_\alpha$ determined by the final BH mass and dimensionless spin, $M_f$ and $\chi_f$.
The natural angular basis for the QNMs is the (spin-weight~${-2}$) spheroidal harmonics ${}_{-2}S_{\alpha}$ which do not align with the ${}_{-2}Y_{\beta}$, leading to mode mixing.
This model applies after the ringdown start time, $t>t_0$.
Throughout, the notation and conventions of Ref.~\cite{2026PhRvD.113j4031D} are used.

This model can be extended by adding an extra damped sinusoid term for the DW in each $\beta$ harmonic;
\begin{align} \label{eq:h_DW}
    rh_{\rm DW}(t,\theta,\phi) \!=\! M \sum_\beta C^\beta_{\mathrm{DW}} e^{-i\omega^\beta_{\mathrm{DW}}(t-t_0)} {}_{-2}Y_{\beta}(\theta,\phi) .
\end{align}
The full ringdown model is $h_{\rm QNM}+h_{\rm DW}$.
The amplitudes $C^\beta_{\mathrm{DW}}$ are free parameters.
If the DW frequencies are set using the horizon properties \cite{2025arXiv250909165O}, as given by Eq.~\ref{eq:DW_freq_beta},
then this is referred to as a horizon mode.
For a Kerr BH, the horizon frequency and surface gravity can be related to the BH mass and spin via
\begin{align} \label{eq:OmegaH_kappa}
    \Omega_{\rm H} = \frac{M_f^{-1}\chi_f}{2\left(1\!+\!\sqrt{1\!-\!\chi_f^2}\right)} , \quad
    \kappa = \frac{M_f^{-1}\sqrt{1-\chi_f^2}}{2\left(1\!+\!\sqrt{1\!-\!\chi_f ^2}\right)} .
\end{align}
We focus on the DW in the dominant $\beta=(2,2)$ harmonic. 
The DW model in Eq.~\ref{eq:h_DW} neglects mode mixing and treats each spherical harmonic independently.

Complex angular frequencies can be written as ${\omega = 2\pi f - i/\tau}$.
Figure \ref{fig:DW_QNM_frequency_comparison} compares the real frequencies and damping times of the horizon mode to other QNMs.
For remnant BHs with spins $\chi_f \approx 0.7$, the ($m=2$) horizon mode has a frequency comparable to the fundamental $\alpha=(2,2,0,+)$ QNM and a damping time in between the fundamental QNM and the first overtone $(2,2,1,+)$.

\begin{figure}[t]
    \includegraphics[width=0.99\linewidth]{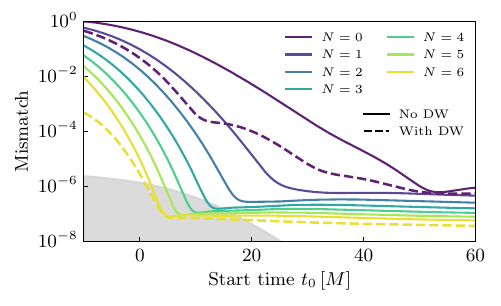}
    \caption{ \label{fig:mismatch}
        Mismatches between the $\beta=(2,2)$ waveform and ringdown models with differing mode content. 
        The colored solid lines show results for QNM overtone models with $n\leq N$.
        The dashed lines show results for models including the DW, the QNM content is indicated by the line color.
        The shaded region is an estimate of the numerical uncertainty obtained by comparing the waveforms at the two highest resolutions.
    }
\end{figure}

\begin{figure*}[t]
    \includegraphics[width=0.99\linewidth]{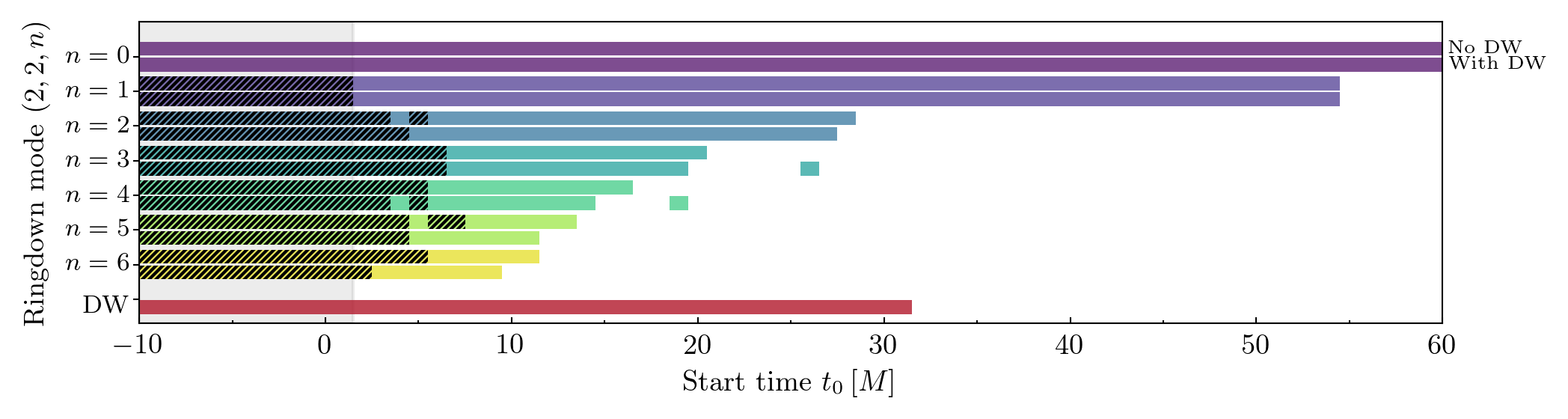}
    \caption{ \label{fig:mode_content}
        The ringdown mode content as a function of ringdown start time.
        The analysis included QNMs with $(\ell,m)=(3,2)$, but results are only shown here for QNMs with  $(\ell,m)=(2,2)$ and the $\beta=(2,2)$ DW.
        A horizontal bar indicates start times are which a mode is confidently identified.
        For each QNM two bars are plotted: for analyses without (top) and with (bottom) the DW included.
        A single horizontal bar is plotted for the DW.
        The fundamental QNM is found at all start times, the first overtone out to $t_0\approx 55M$, the DW and the second overtone out to $t_0\approx 30M$, and the higher overtones only at earlier start times (as expected due to their faster decay rates).
        Retrograde QNMs, indicated by hatched bars, are only found at early start times, usually where the model gives a poor fit to the NR data as judged by the PPC (shaded gray region).
        In a few cases a mode that has dropped out of the model is briefly identified again at later times; this numerical artifact was previously described in Ref.~\cite{2026PhRvL.136s1403D}.
        This QNM content is consistent with that found by a similar analysis in Ref.~\cite{2026PhRvL.136s1403D} without the DW.
    }
\end{figure*}

{\textit{\textbf{Results}}---}In this Letter we use NR waveforms from the public CCE catalog \cite{SXS_CCE_catalog}.
These are preprocessed following Ref.~\cite{2026PhRvD.113j4031D} to transform to the superrest frame and align the peak of the $\beta=(2,2)$ harmonic strain at $t=0$.
The main text shows results for the equal-mass, high-spin, non-precessing simulation CCE:0004 which shows a DW most clearly; results for other simulations are provided in the Supplemental Material.

The quality of fit of a ringdown model can be judged using a simple least-squares fit to the dominant $\beta=(2,2)$ harmonic. 
A QNM overtone model including prograde ${\ell=m=2}$ QNMs and with $n\leq N$ was used for a range of start times both with and without the DW included.
The results are shown in Fig.~\ref{fig:mismatch} using the mismatch (defined, for example, in Eq.~6 of Ref.~\cite{2025PhRvD.111b4002D}).
Including the DW improves the mismatch, particularly at early times, compared to a similar QNM-only model.
By itself, this is not evidence for the DW due to the possibility of overfitting when using least-squares and the differing numbers of parameters in the ringdown models used.

To overcome these difficulties we use the Bayesian method introduced in Ref.~\cite{2026PhRvD.113j4031D} for QNMs. 
This uses a Gaussian process (GP) model for the uncertainty in the NR waveforms trained on the residuals between different numerical resolutions for all simulations in the catalog.
All of the fits performed here were done using the Bondi news, $\mathcal{N}^{\beta}(t)=\dot{\mathfrak{h}}^\beta(t)$.
The model is a sum of damped sinusoids with frequencies fixed to the predicted values (either QNM or DW) calculated using the remnant mass and spin.
The free parameters are the amplitudes $C_\alpha$ with flat, unbounded priors on the real and imaginary parts.
Because the model depends linearly on these parameters, the posterior is a multivariate Gaussian and can be sampled extremely efficiently.

The evidence ratio (or Bayes factor) between ringdown models with differing numbers of modes was used in Ref.~\cite{2026PhRvL.136s1403D} to define a Bayesian method for determining the ringdown mode content. 
Modes are added sequentially to the model in the order determined by their Bayesian evidence, terminating when the evidence for any new modes drops below a specified threshold. 
This procedure is applied separately at different start times allowing a picture of the evolving mode content to be produced. 
A posterior predictive check (PPC) is used to judge when the resulting model gives an adequate fit to the NR waveform.
For full details of the algorithm, see Ref.~\cite{2026PhRvL.136s1403D}.
Here this is extended to search for the DW signal by including $h_{\rm DW}$ in the list of candidate modes checked at each iteration.

This was used to search for the DW in the $\beta=(2,2)$ harmonic.
The waveform harmonics $\beta=(2,2)$ and $(3,2)$ were fit (allowing for QNM mode mixing) using the candidate QNMs $\alpha=(2,2,n,p)$ and $(3,2,n,p)$ with $n\leq 6$ and $p=\pm$.
Constant offset terms were also included.
Runs were performed both with and without the $\beta=(2,2)$ DW.
The evolving mode content is plotted in Fig.~\ref{fig:mode_content}.

The DW is confidently identified (with a significance $\gtrsim 4\sigma$, see Ref.~\cite{2026PhRvL.136s1403D}) for start times up to $t_0 \approx 31M$ after the peak strain.
Including the DW does not significantly affect the other QNMs; both with and without the DW, the fundamental and first overtone are identified over identical times ranges, and only small differences of $\lesssim 2M$ are seen for higher overtones ($n\geq 2$).

The Bayesian analysis method also produces posteriors for all of the mode amplitudes in the model. 
The DW has an amplitude approximately a factor of ten below that of the fundamental QNM and has a stable value for a wide range of start times.
The amplitudes of the low-order ($n\leq 2$) overtones are unaffected by the inclusion of the DW and only minor effects are seen for high overtones.
The amplitude posteriors are plotted in Fig.~\ref{fig:mode_amplitudes}.

\begin{figure*}[t]
    \includegraphics[width=0.99\linewidth]{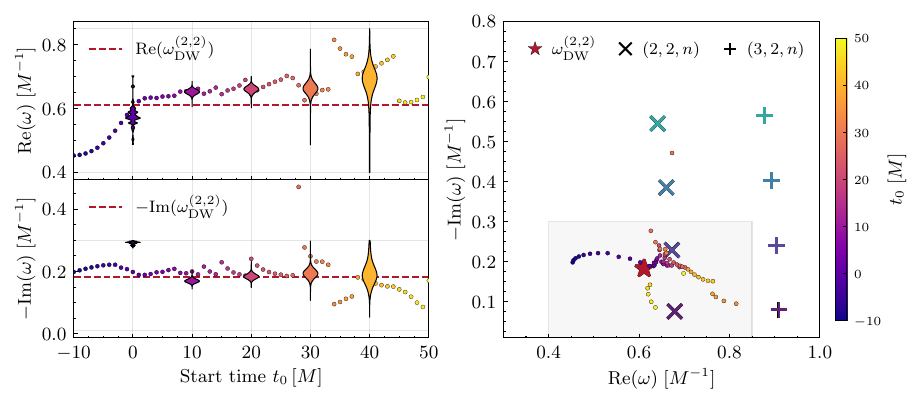}
    \caption{ \label{fig:free_frequency}
        The results of the free frequency fits. 
        Violin plots in the left panels show the posteriors on the real and imaginary parts of the DW frequency, $\omega_{\rm DW}$.
        Horizontal dashed lines show the predicted horizon mode values.
        Dots show the frequencies found by non-linear least-squares fits across the full range of start times.
        The discontinuities are associated with QNMs dropping out of the model; see Fig.~\ref{fig:mode_content}.
        The least-squares results are shown as a complex frequency in the right panel where they are compared with the horizon mode (star) and select QNM frequencies (plus and cross).
        The gray box shows the $\omega$ prior range.
    }
\end{figure*}

The mode content calculations described above were done with the DW frequency fixed to the horizon mode value in Eq.~\ref{eq:DW_freq_beta}.
Another way of confirming the presence of a ringdown mode is to perform a fit where the mode frequency is allowed to vary as a (non-linear) free parameter.
Both amplitude stability and free frequency fits have been widely used to confirm the presence of QNMs in the ringdown (see, for example, Ref.~\cite{Giesler:2024hcr}).
The Bayesian analysis method can be adapted to perform free frequency fits \cite{Dyer_in_prep}.
At each start time the list of QNMs to be included was set to the mode content determined in Fig.~\ref{fig:mode_content} with the $\beta=(2,2)$ DW included.
All the QNM frequencies were fixed, but the DW frequency was allowed to vary with flat priors on both the real and imaginary parts over the ranges indicated in Fig.~\ref{fig:free_frequency}.

The posteriors on the frequency and damping time are plotted in the left panel of Fig.~\ref{fig:free_frequency}.
These non-linear models have non-Gaussian posteriors that are sampled using more expensive Markov chain Monte Carlo methods; therefore, these analyses were only done at start times $t_0/M=0,\,10,\,20,\,30,\,40$.
Intermediate start times were filled in using the results of a quicker non-linear least-squares fit.
Least-squares fits are equivalent to maximum likelihood estimates using a white-noise model and are therefore not expected to exactly coincide with the peak of the Bayesian posteriors obtained using the GP noise model.
The results of the least-squares fits are plotted in both panels of Fig.~\ref{fig:free_frequency} using dots colored based on the ringdown start time.
These results show that while the frequency of the DW is close to the horizon mode prediction for $t_0 > 0M$, its precise value drifts by $\sim 10-20\%$ over time and doesn't track the horizon-mode frequency, broadly consistent with Ref.~\cite{Kankani:2026byb}.
To some extent, this is expected on physical grounds as (in the point-particle picture) the DW should only tend to have the horizon mode frequency as the plunging particle approaches the horizon.
We also note that the recovered frequency of the DW can depend sensitively on which other QNMs are included in the model.

\begin{figure}[t]
    \includegraphics[width=0.99\linewidth]{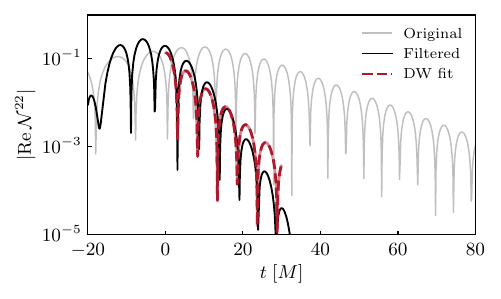}
    \caption{ \label{fig:rational_filter}
        Rational filter results. 
        The original waveform (gray) is filtered to remove three of the loudest QNMs (black). 
        This filtered signal is fit over a range of times using a single damped sinusoid with the horizon mode frequency, the dashed line (red) and shaded region show the median and 50\% credible interval of the fit.
        This gives a reasonable fit to the early time ringdown, confirming the presence of the DW.
        The fit is not expected to be perfect due to presence of other (unfiltered) QNMs and the fact that the actual DW frequency deviates from frequency predicted by the horizon mode (see Fig.~\ref{fig:free_frequency}).
    }
\end{figure}

The ringdown mode content can also be analyzed using rational filters \cite{Lu:2025mwp, Ma:2022wpv}. 
This approach attempts to remove modes with a known frequency from a time series without having to fit for their amplitudes or specify a start time.
Here the $\alpha=(2,2,0,+)$, $(2,2,1,+)$, and $(3,2,0,+)$ QNMs were filtered out of the $\beta=(2,2)$ waveform. 
Once the known, loud modes have been removed, quieter subdominant modes are easier to identify in the filtered signal.
The DW was searched for by fitting a single mode with the horizon frequency over the period $0\leq t/M\leq 30$; the resulting fit is shown in Fig.~\ref{fig:rational_filter} and should be compared to Fig.~3 in Ref.~\cite{2025arXiv251001001L}.
This provides an independent confirmation of the presence of the DW.

The excitation amplitudes of the QNMs are known to depend on the parameters of the progenitor binary (see e.g.\ Refs.~\cite{2024PhRvD.110j3037P, Mitman:2025hgy}). 
The same is expected to be true for the DW part of the ringdown.
Nevertheless, the results found here for the $\beta=(2,2)$ DW in simulation CCE:0004 are representative of those found in other simulations in the public CCE catalog.
Bayesian mode content calculations including the DW have been performed for these simulations and the results are summarized in Tab.~\ref{tab:catalog}.
The DW has also been searched for and confidently identified in other harmonics besides $\beta=(2,2)$; results for the $\beta=(3,3)$ DW in CCE:0010 are shown in Fig.~\ref{fig:DW_harmonic}.

{\textit{\textbf{Discussion}}---}It has been suggested that the DW part of the ringdown can be simply modeled as a damped sinusoid with a frequency related to properties of the event horizon.
Here, this has been tested using state-of-the-art NR waveforms. 
Techniques originally designed for QNMs have been adapted to analyze DWs, including a robust Bayesian framework for detecting them.
Modeling the DW using a horizon mode improves the agreement with the NR waveforms and the inclusion of the DW is favored by the Bayesian evidence for a wide range of start times up to $30M$ after the peak of the strain. 

We find the DW with a stable amplitude slightly below that of the fundamental QNM.
However, when the frequency of the DW is allowed to vary, it does not reliably recover the predicted horizon mode.
This is partially expected given the physical picture of the DW being sourced by a plunging particle as it is seen to approach (but never reach) the horizon.
These results suggest that the DW can be modeled using a horizon mode and can be usefully incorporated into ringdown models for GW data analysis, as done in \cite{2025arXiv251001001L, 2026arXiv260606592K}.
However, these results call into question claims that the results can be used to directly probe the geometry near the horizon.

{\textit{\textbf{Acknowledgments}}---}
RD acknowledges support from the STFC for their PhD (Grant No.\ ST/Y509139/1) and AC acknowledges the Herchel Smith Research Fellowship.

{\textit{\textbf{Data Availability}}---}The code to perform the Bayesian ringdown fits is available at \href{https://github.com/BGP-QNM-FITS/bgp_qnm_fits}{\faGithub}~\cite{bgp_qnm_fits} and all analysis and plotting scripts are available at \href{https://github.com/Richardvnd/direct_waves}{\faGithub}~\cite{DW_fits}. 
We are grateful to the SXS collaboration for making the NR CCE waveform data publicly available \cite{SXS_CCE_catalog}.

\vspace{1cm}
\bibliographystyle{apsrev4-1}
\bibliography{references}

\clearpage
\onecolumngrid
\newpage
\begin{center}
  \textbf{\large{Supplemental Material}} \\
\end{center}
\twocolumngrid

\setcounter{equation}{0}
\setcounter{figure}{0}
\setcounter{table}{0}
\setcounter{page}{1}
\makeatletter
\renewcommand{\theequation}{S\arabic{equation}}
\renewcommand{\thefigure}{S\arabic{figure}}
\renewcommand{\thetable}{S\arabic{figure}}

\section*{Mode amplitudes}
\begin{figure}[h]
    \includegraphics[width=0.99\linewidth]{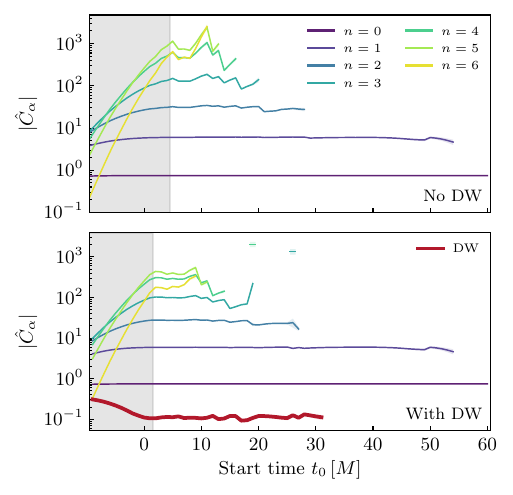}
    \caption{ \label{fig:mode_amplitudes}
        The ringdown mode amplitudes as functions of start time.
        The bottom (top) panel shows the results from Bayesian analyses performed with (without) the DW included.
        The median and 50\% credible posterior interval on its amplitude is plotted; many of these credible intervals are too narrow to be seen on this scale.
        The gray shaded regions show where the PPC indicates the ringdown model gives a poor fit to the data; the region in the top panel coincides with the one shown in Fig.~\ref{fig:mode_content}.
    }
\end{figure}

The Bayesian fits used to determine the ringdown mode content in the main Letter also produce posteriors on all of the mode amplitudes. 
For all times when a particular mode is identified the posterior on its mode amplitude $|\hat{C}_\alpha|$ is plotted in Fig.~\ref{fig:mode_amplitudes}.
The hat denotes that these amplitudes have been decay-corrected by multiplying the recovered amplitude $|C_\alpha|$ by an exponential designed to remove the expected exponential decay; a flat line in the figure indicates a mode decaying at the expected rate.
For QNMs that drop out of the model before briefly reappearing again at later start times, the amplitudes are observed to change significantly; this is the origin of the very short, high-amplitude line segments in the bottom panel of the figure.

\vspace{2cm}
\section*{Catalog Results}
For each of the 13 simulations in the public CCE waveform catalog a Bayesian QNM mode content calculation was performed, similar to that shown in Fig.~\ref{fig:mode_content} for CCE:0004.
Together, these 13 simulations cover a range of mass ratios, component spin magnitudes and spin orientations in the binary BH parameter space and they allow us to investigate how this affects the DW.
These mode content calculations focused on the $\beta=(2,2)$ harmonic and the DW was clearly identified in the first 9 catalog simulations and somewhat less clearly in simulations CCE:0010-0012.
The final, highly-asymmetric binary in simulation CCE:0013 is omitted because it is known to yield very atypical ringdown fits (see, for example, Ref.~\cite{2026PhRvL.136s1403D}).
Table \ref{tab:catalog} reports the amplitude of DWs in the $\beta=(2,2)$ harmonics for each simulation along with the range of start times at which the DW was confidently identified and mode content plots are available at \href{https://github.com/Richardvnd/direct_waves}{\faGithub}~\cite{DW_fits}. 

\begin{table}[h]
    \caption{ \label{tab:catalog}
        For each simulation in the catalog, this table reports the amplitude of the $\beta=(2,2)$ DW as a ratio of the amplitude of the fundamental QNM at a fixed start time of $t_0=5M$ as well as the range of start times $t_0<t_{0,\,\mathrm{max}}$ for which the DW was reliably identified. The mean and $\pm 1\sigma$ error bars were estimated from the posterior samples.
        Rows marked $^*$ indicate the DW disappears and reappears such that $t_{0,\,\mathrm{max}}$ is unclear.
        Mode content plots are available at \href{https://github.com/Richardvnd/direct_waves}{\faGithub}~\cite{DW_fits}.
    }
\begin{ruledtabular}
\begin{tabular}{lcr}
    Sim ID & $|C^{(2,2)}_{\rm DW}|/|C_{(2,2,0,+)}|$ & $t_{0,\,\mathrm{max}}$ $[M]$ \\
    \hline
    0001 & $0.09 \pm 0.02$ & 12 \\
    0002 & $0.12 \pm 0.01$ & 13 \\
    0003 & $0.15 \pm 0.01$ & 18 \\
    0004 & $0.151 \pm 0.007$ & 31 \\
    0005 & $0.09 \pm 0.03$ & 11 \\
    0006 & $0.11 \pm 0.02$ & 12* \\
    0007 & $0.25 \pm 0.03$ & 11 \\
    0008 & $0.26 \pm 0.02$ & 9 \\
    0009 & $2.10 \pm 0.05$ & 25 \\
    0010 & $0.45 \pm 0.05$ & 6 \\
    0011 & $0.41 \pm 0.04$ & 11* \\
    0012 & $0.77 \pm 0.03$ & 25* \\
\end{tabular}
\end{ruledtabular}
\end{table}

\section*{Direct Wave Harmonics}
\begin{figure*}[t]
    \includegraphics[width=0.99\linewidth]{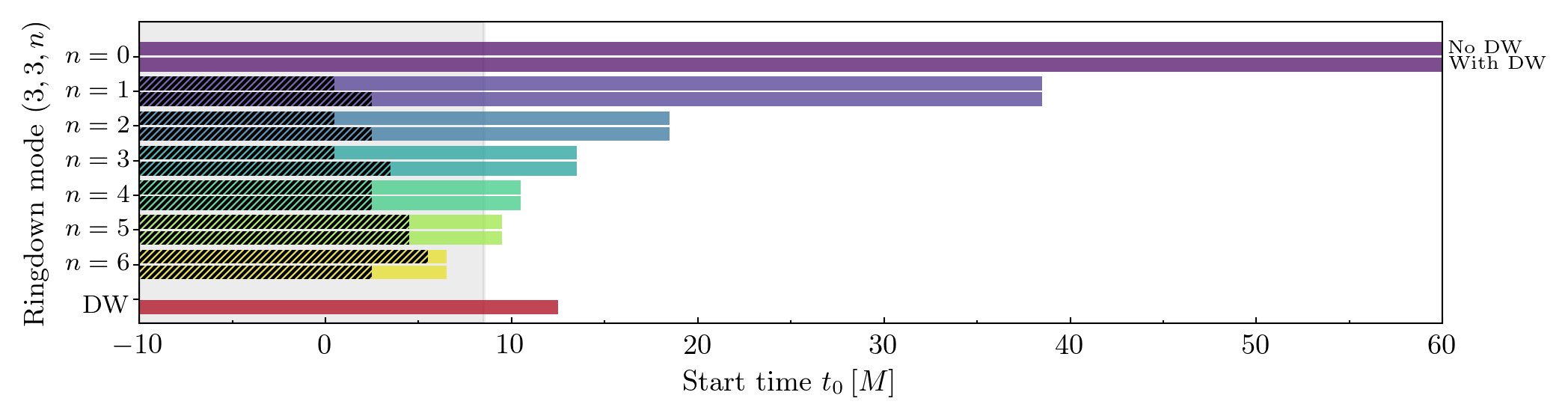}
    \caption{ \label{fig:DW_harmonic}
        Similar to Fig.~\ref{fig:mode_content}, the $\beta=(3,3)$ ringdown mode content as a function of start time for simulation CCE:0010.
        The analysis included QNMs with $(\ell,m)=(4,3)$, but mode content results are only shown here for QNMs with  $(\ell,m)=(3,3)$ and the $\beta=(3,3)$ DW.
        A horizontal bar indicates ringdown start times at which a mode is confidently identified.
        For each QNM two bars are plotted for the analyses without (top) and with (bottom) the DW.
        A single horizontal bar is plotted for the DW.
        The $\beta=(3,3)$ DW is confidently identified at early times, but this only extends out to $t_0\approx 13M$ after the peak strain.
    }
\end{figure*}

The main text focuses on the DW in the dominant $\beta=(2,2)$ waveform harmonic.
DWs are also expected in the subdominant, higher harmonics. 
Figure \ref{fig:DW_harmonic} shows the results of a search for the DW in the $\beta=(3,3)$ harmonic of the non-spinning, mass ratio 4:1 simulation CCE:0010.

\end{document}